\documentclass[prd,onecolumn,nofootinbib,showpacs,showkeys]{revtex4}

\begin{document}

%\markboth{N. J. Pop\l awski}
%{On the Nonsymmetric Purely Affine Gravity}

%%%%%%%%%%%%%%%%%%%%% Publisher's Area please ignore %%%%%%%%%%%%%%
%\catchline{}{}{}{}{}
%\catchline{Vol. 22, No. 36 (2007) 2701--2720}
%%%%%%%%%%%%%%%%%%%%%%%%%%%%%%%%%%%%%%%%%%%%%%%%%%%%%%%%%%%%%%%%%%%

\title{ON THE NONSYMMETRIC PURELY AFFINE GRAVITY}

\author{{\bf Nikodem J. Pop\l awski}}

\affiliation{Department of Physics, Indiana University, Swain Hall West, 727 East Third Street, Bloomington, IN 47405, USA}
\email{nipoplaw@indiana.edu}

\noindent
%Electronic version of an article published as:
{\em Modern Physics Letters A}\\
Vol. {\bf 22}, No. 36 (2007) 2701--2720\\
%\htmladdnormallink{DOI: 10.1142/S0217732307025662}{http://dx.doi.org/10.1142/S0217732307025662}\\
\copyright\,World Scientific Publishing Co.
\vspace{0.4in}

%\pub{Received (Day Month Year)}{Revised (Day Month Year)}
%\pub{Vol. 22, No. 36 (2007) 2701--2720}

\begin{abstract}
We review the vacuum purely affine gravity with the nonsymmetric connection and metric.
We also examine dynamical effects of the second Ricci tensor and covariant second-rank tensors constructed from the torsion tensor in the gravitational Lagrangian.
\end{abstract}


\maketitle

\section{Introduction}
\label{secIntro}

In the {\em purely affine} (Einstein--Eddington) formulation of general relativity~\cite{Ein1923a,Ein1923b,Edd,Sch1947a,Sch1950,Kij,CFK,KW}, a Lagrangian density depends on a torsionless affine connection and the symmetric part of the Ricci tensor of the connection.
This formulation defines the metric tensor as the derivative of the Lagrangian density with respect to the Ricci tensor, obtaining an algebraic relation between these two tensors.
It derives the field equations by varying the total action with respect to the connection, which gives a differential relation between the connection and the metric tensor.
This relation yields a differential equation for the metric.
In the {\em metric--affine} (Einstein--Palatini) formulation~\cite{Ein1923a,Pal,Lord,FFR}, both the metric tensor and the torsionless connection are independent variables, and the field equations are derived by varying the action with respect to these quantities.
The corresponding Lagrangian density is linear in the symmetric part of the Ricci tensor of the connection.
In the {\em purely metric} (Einstein--Hilbert) formulation~\cite{Ein1916a,Hilb,Lor1916a,Lor1916b,Lor1916c,Lor1916d,LL}, the metric tensor is a variable, the affine connection is the Levi-Civita connection of the metric and the field equations are derived by varying the action with respect to the metric tensor.
The corresponding Lagrangian density is linear in the symmetric part of the Ricci tensor of the metric.
All three formulations of general relativity are dynamically equivalent~\cite{FK1982a}.
This statement can be generalized to theories of gravitation with Lagrangians that depend on the full Ricci tensor and the second Ricci tensor~\cite{JK1989a,JK1989b}, and to a general connection with torsion~\cite{JK1989b}.
There also exist formulations of gravity in which the dynamical variables are: metric and torsion (Einstein--Cartan theory)~\cite{Car1,Car2,Car3,Car4,Hehl,SG}, metric and nonsymmetric connection~\cite{HK,HLS}, tetrad and spin connection (Einstein--Cartan--Kibble--Sciama theory)~\cite{Uti,Kib,Sci1,Sci2,Cho,VCB,BC}, spin connection~\cite{CJD,CDJ,CJ}, and spinors~\cite{TJ,Wet}.

The fact that Einstein's relativistic theory of gravitation~\cite{Ein1916b} is based on the affine connection rather than the metric tensor was first noticed by Weyl~\cite{Weyl}.
Shortly after, this idea was developed by Eddington, who constructed the simplest purely affine gravitational Lagrangian~\cite{Edd}.
In a series of beautiful papers, entitled {\em The Final Affine Field Laws}~\cite{Sch1947b,Sch1948a,Sch1948b}, Schr\"{o}dinger elucidated Eddington's affine theory and generalized it to a nonsymmetric connection, introduced earlier by Cartan~\cite{Car1,Car2,Car3,Car4,Hehl}.
Schr\"{o}dinger's affine theory~\cite{Sch1943,Sch1944a,Sch1944b,Sch1946,Sch1951} used a nonsymmetric metric tensor, which was introduced earlier by Einstein and Straus~\cite{Ein1925,Ein1945,ES,Ein1948,Ein1950} in the metric formulation to unify gravitation with electromagnetism (Einstein's unified field theory~\cite{Hla1952,Hla1953,HS,Hla1954a,Hla1954b,Hla1954c,Hla1957,Ton,Goe}).
The antisymmetric part of this tensor was supposed to correspond to the electromagnetic tensor~\cite{Sch1948a,Sch1948b,Mof1975}.
The failure of such an association~\cite{Inf,Kur1,Cal,Kur2} resulted in the nonsymmetric gravitational theory of Moffat~\cite{Mof1979,Mof1987}, according to which the nonsymmetric metric tensor describes only gravity. 
As the Einstein--Straus theory, Moffat's theory uses the metric--affine variational principle.

In the Eddington--Schr\"{o}dinger purely affine formulation of gravity, the connection is the fundamental variable, analogous to the coordinates in relativistic mechanics.
The relation between the purely affine and metric--affine formulation of gravity is analogous to the relation between Lagrangian and Hamiltonian dynamics in classical mechanics~\cite{FK1982a}.
The curvature corresponds to the four-dimensional velocities, and the metric corresponds to the generalized momenta~\cite{Kij}.
Remarkably, this analogy can be completed by noticing that both the connection and the coordinates are not tensors, but their variations are (the coordinate variations are vectors by definition).
The curvature and metric form tensors and so do the velocities and momenta.
The metric--affine Lagrangian density for the gravitational field is precisely the Legendre term in the Legendre transformation with respect to the Ricci tensor, applied to a purely affine Lagrangian.
This term automatically turns out to be linear in the curvature scalar~\cite{FK1982a}, distinguishing (with respect to the purely affine picture) general relativity and its nonsymmetric extensions from nonlinear theories of gravity with metric structure.
The purely affine formulation differs from the Einstein--Straus theory because it automatically generates the cosmological term in the field equations.

In this paper we review the nonsymmetric purely affine gravity in vacuum and examine the role of the curvature in the field dynamics.
In Sec.~\ref{secLagr} we introduce the spacetime structure in the purely affine gravity.
In Sec.~\ref{secField} we derive the affine field equations, and in Sec.~\ref{secMot} we discuss the equations of motion on geodesics.
Sec.~\ref{secSol} examines the spherically symmetric solution for the purely affine gravity in vacuum.
In Sec.~\ref{secRic} we study the role of the second Ricci tensor in the affine gravitational Lagrangian, and in Sec.~\ref{secTor} we consider the effects of the torsion tensor in this Lagrangian on the field equations.
We briefly summarize the results in Sec.~\ref{secSum}.

\section{The Eddington--Schr\"{o}dinger Lagrangian}
\label{secLagr}

A contravariant vector $V^\mu$ and a covariant vector $V_\mu$ are defined as the quantities which
transform under general coordinate transformations like a coordinate differential and a gradient, respectively.
The differentials $dV^\mu$ and $dV_\mu$ are not vectors with respect to these transformations, and need to be replaced by the covariant differentials:
\begin{eqnarray}
{\cal D}V^\mu=dV^\mu-\delta V^\mu, \label{vec1}\\
{\cal D}V_\mu=dV_\mu-\delta V_\mu, \label{vec2}
\end{eqnarray}
where $\delta$ denotes an infinitesimal parallel translation.
The concept of a parallel translation of a vector introduces the linear affine connection $\Gamma^{\,\,\rho}_{\mu\,\nu}$:
\begin{eqnarray}
& & \delta V^\mu=-\Gamma^{\,\,\mu}_{\rho\,\nu}V^\rho dx^\nu, \label{con1}\\
& & \delta V_\mu=\Gamma^{\,\,\rho}_{\mu\,\nu}V_\rho dx^\nu, \label{con2}
\end{eqnarray}
which enters the definition of the covariant derivative of a vector:
\begin{eqnarray}
& & V^\mu_{\phantom{\mu};\nu}=V^\mu_{\phantom{\mu},\nu}+\Gamma^{\,\,\mu}_{\rho\,\nu}V^\rho, \label{cov1}\\
& & V_{\mu;\nu}=V_{\mu,\nu}-\Gamma^{\,\,\rho}_{\mu\,\nu}V_\rho, \label{cov2}
\end{eqnarray}
where the comma denotes the usual partial derivative.
The order of the two covariant indices in the connection coefficients is important since we do not restrict the connection to be symmetric~\cite{Ein1925,Ein1945,ES,Ein1948,Ein1950}.
The covariant derivative is assumed to obey the chain rule of differentiating products, and the transformation law for an arbitrary tensor results from the definition of a tensor as the quantity which transforms like the corresponding product of vectors.
 
The purely affine variational principle regards the 64 components of $\Gamma^{\,\,\rho}_{\mu\,\nu}$ as the gravitational field variables~\cite{Sch1950}.
Under inhomogeneous coordinate transformations $x^\mu\rightarrow x^{\mu'}(x^\mu)$ the connection transforms according to:
\begin{equation}
\Gamma^{\,\,\,\rho'}_{\mu'\,\nu'}=\frac{\partial x^{\rho'}}{\partial x^\rho}\frac{\partial x^\mu}{\partial x^{\mu'}}\frac{\partial x^\nu}{\partial x^{\nu'}}\,\Gamma^{\,\,\rho}_{\mu\,\nu}+\frac{\partial x^{\rho'}}{\partial x^\rho}\frac{\partial^2x^{\rho}}{\partial x^{\mu'}\,\partial x^{\nu'}}.
\label{con3}
\end{equation}
It follows that the connection is not a tensor, unlike its antisymmetric part which is the Cartan torsion tensor,
\begin{equation}
S^\rho_{\phantom{\rho}\mu\nu}=\Gamma^{\,\,\,\,\rho}_{[\mu\,\nu]}.
\label{tor1}
\end{equation}
The contraction of the torsion tensor defines the torsion vector,
\begin{equation}
S_\mu=S^\nu_{\phantom{\nu}\mu\nu}.
\label{tor2}
\end{equation}
We also introduce the modified connection (Schr\"{o}dinger's star-affinity)~\cite{Sch1950,Sch1947b}
\begin{equation}
^\ast\Gamma^{\,\,\rho}_{\mu\,\nu}=\Gamma^{\,\,\rho}_{\mu\,\nu}+\frac{2}{3}\delta^\rho_\mu S_\nu,
\label{con4}
\end{equation}
which obeys 
\begin{equation}
^\ast S_\mu=\,^\ast\Gamma^{\,\,\,\,\nu}_{[\mu\,\nu]}=0.
\label{con5}
\end{equation}

As in general relativity, the curvature tensor
\begin{equation}
R^\rho_{\phantom{\rho}\mu\sigma\nu}=\Gamma^{\,\,\rho}_{\mu\,\nu,\sigma}-\Gamma^{\,\,\rho}_{\mu\,\sigma,\nu}+\Gamma^{\,\,\kappa}_{\mu\,\nu}\Gamma^{\,\,\rho}_{\kappa\,\sigma}-\Gamma^{\,\,\kappa}_{\mu\,\sigma}\Gamma^{\,\,\rho}_{\kappa\,\nu},
\label{curv1}
\end{equation}
is defined through a parallel displacement of a vector $V_\mu$ along the boundary of an infinitesimal surface element $\Delta f^{\mu\nu}$~\cite{Sch1950}:
\begin{equation}
\oint\delta V_\rho=\frac{1}{2}R^\sigma_{\phantom{\sigma}\rho\mu\nu}V_\sigma\Delta f^{\mu\nu},
\label{disp}
\end{equation}
or by the commutator of the covariant derivatives:
\begin{equation}
V^\rho_{\phantom{\rho};\nu\mu}-V^\rho_{\phantom{\rho};\mu\nu}=R^\rho_{\phantom{\rho}\sigma\mu\nu}V^\sigma+2S^\sigma_{\phantom{\sigma}\mu\nu}V^\rho_{\phantom{\rho};\sigma}.
\label{curv2}
\end{equation}
For a nonsymmetric connection, the curvature tensor is not antisymmetric in its first two indices which results in two possibilities of contraction~\cite{Edd,Scho}.
The usual Ricci tensor is defined as
\begin{equation}
R_{\mu\nu}=R^\rho_{\phantom{\rho}\mu\rho\nu},
\label{Ric1}
\end{equation}
which gives
\begin{equation}
R_{\mu\nu}=\Gamma^{\,\,\rho}_{\mu\,\nu,\rho}-\Gamma^{\,\,\rho}_{\mu\,\rho,\nu}+\Gamma^{\,\,\kappa}_{\mu\,\nu}\Gamma^{\,\,\rho}_{\kappa\,\rho}-\Gamma^{\,\,\kappa}_{\mu\,\rho}\Gamma^{\,\,\rho}_{\kappa\,\nu}.
\label{Ric2}
\end{equation}
For a general connection, the tensor~(\ref{Ric2}) is not symmetric.
The second Ricci tensor is defined as
\begin{equation}
Q_{\mu\nu}=R^\rho_{\phantom{\rho}\rho\mu\nu},
\label{Ric3}
\end{equation}
from which it follows that this tensor is antisymmetric and has the form of a curl:
\begin{equation}
Q_{\mu\nu}=\Gamma^{\,\,\rho}_{\rho\,\nu,\mu}-\Gamma^{\,\,\rho}_{\rho\,\mu,\nu}.
\label{Ric4}
\end{equation}
In general relativity, this tensor vanishes due to the symmetries of the Riemann curvature tensor.

The condition for a Lagrangian density to be covariant is that it must be a product of a scalar and the square root of the determinant of a covariant tensor of rank two~\cite{Sch1950}.
The simplest curvature-only dependent Lagrangian density of this form was introduced by Eddington~\cite{Edd} and generalized by Schr\"{o}dinger~\cite{Sch1950}:\footnote{
Eddington used the symmetric connection $\Gamma^{\,\,\rho}_{\mu\,\nu}=\Gamma^{\,\,\,\,\rho}_{(\mu\,\nu)}$, for which the tensors $R_{\mu\nu}$ and $Q_{\mu\nu}$ are not independent since $Q_{\mu\nu}=2R_{[\mu\nu]}$, and the Lagrangian density ${\cal L}=-\frac{2}{\Lambda}\sqrt{-\mbox{det}R_{(\mu\nu)}}$.}\footnote{
The quantity $d^4x\sqrt{-\mbox{det}R_{(\mu\nu)}}$ is referred to by Eddington as the generalized volume element~\cite{Edd}.}
\begin{equation}
{\cal L}=-\frac{2}{\Lambda}\sqrt{-\mbox{det}R_{\mu\nu}},
\label{Lagr1}
\end{equation}
where $\Lambda$ is a constant.
Consequently, the gravitational action is given by\footnote{
We set the units so that $c=1$.}
\begin{equation}
S=\int d^4x{\cal L}.
\label{action}
\end{equation}  
The metric structure associated with this Lagrangian is obtained using the following prescription~\cite{Edd,Sch1950,Sch1947b}:
\begin{equation}
{\sf g}^{\mu\nu}=\frac{\partial{\cal L}}{\partial R_{\mu\nu}},
\label{met1}
\end{equation}
where ${\sf g}^{\mu\nu}$ is the fundamental tensor density.\footnote{
If $\Lambda$ has the dimension of $\textrm{m}^{-2}$ (cosmological constant), the correct dimension of the action can be obtained by dividing the right-hand side of Eq.~(\ref{Lagr1}) by $-2\kappa$, where $\kappa$ is Einstein's gravitational constant, and multiplying the right-hand side of Eq.~(\ref{met1}) by $-2\kappa$~\cite{Kij,FK1982a,FK1981}.
However, since we consider the gravitational field in vacuum, the factor $-2\kappa$ is irrelevant.
Moreover, we can always set the units so that $-2\kappa=1$.}
The contravariant metric tensor is defined by~\cite{Sch1950,Pap}
\begin{equation}
g^{\mu\nu}=\frac{{\sf g}^{\mu\nu}}{\sqrt{-\mbox{det}{\sf g}^{\rho\sigma}}}.
\label{met2}
\end{equation}
To make this definition meaningful, we must assume
\begin{equation}
\mbox{det}{\sf g}^{\mu\nu}\neq0.
\label{phys}
\end{equation}
The covariant metric tensor $g_{\mu\nu}$ is related to the contravariant metric tensor by
\begin{equation}
g^{\mu\nu}g_{\rho\nu}=g^{\nu\mu}g_{\nu\rho}=\delta^\mu_\rho,
\label{met3}
\end{equation}
where the order of the indices is important since both metric tensors are nonsymmetric.
Accordingly, there is no general scheme for raising and lowering indices.

Substituting Eq.~(\ref{Lagr1}) into~(\ref{met1}) yields~\cite{Sch1947b}
\begin{equation}
\sqrt{-\mbox{det}R_{\rho\sigma}}K^{\mu\nu}=-\Lambda{\sf g}^{\mu\nu},
\label{cosm1}
\end{equation}
where the tensor $K^{\mu\nu}$ is reciprocal to the Ricci tensor:\footnote{
We use the identity $\delta\sqrt{-\mbox{det}R_{\rho\sigma}}=\frac{1}{2}\sqrt{-\mbox{det}R_{\rho\sigma}}K^{\mu\nu}\delta R_{\mu\nu}$.}
\begin{equation}
K^{\mu\nu}R_{\rho\nu}=K^{\nu\mu}R_{\nu\rho}=\delta^\mu_\rho.
\label{cosm2}
\end{equation}
Eq.~(\ref{cosm1}) is equivalent to
\begin{equation}
R_{\mu\nu}=-\Lambda g_{\mu\nu},
\label{cosm3}
\end{equation} 
which formally has the form of the Einstein field equations of general relativity with the cosmological constant $\Lambda$~\cite{Sch1950}.
In order to find explicit partial differential equations for the metric tensor, we need a relation between $g_{\mu\nu}$ and $\Gamma^{\,\,\rho}_{\mu\,\nu}$.

\section{The Field Equations}
\label{secField}

The dynamics of the gravitational field in the nonsymmetric gravity is governed by the principle of least action $\delta S=0$.
The variation of the action corresponding to the Lagrangian~(\ref{Lagr1}) reads~\cite{Sch1947b}
\begin{equation}
\delta S=\int d^4x\frac{\partial{\cal L}}{\partial R_{\mu\nu}}\delta R_{\mu\nu}=\int d^4x\,{\sf g}^{\mu\nu}\delta R_{\mu\nu},
\label{act2}
\end{equation}
where we vary the action with respect to the connection through $R_{\mu\nu}$.\footnote{
In the Einstein--Straus theory, the gravitational Lagrangian density is given by ${\cal L}={\sf g}^{\mu\nu}R_{\mu\nu}$, where ${\sf g}^{\mu\nu}$ is a quantity independent of $R_{\mu\nu}$.
Consequently, $\delta S=\int d^4x\,{\sf g}^{\mu\nu}\delta R_{\mu\nu}+\int d^4x\,R_{\mu\nu}\delta {\sf g}^{\mu\nu}$, and the variation with respect to ${\sf g}^{\mu\nu}$ leads to the equation $R_{\mu\nu}=0$ which does not contain the cosmological constant~\cite{ES,Ein1948,Ein1950}.}
The variation of the Ricci tensor is given by the Palatini formula~\cite{Sch1950,Pal,Sch1947b},
\begin{equation}
\delta R_{\mu\nu}=\delta\Gamma^{\,\,\rho}_{\mu\,\nu;\rho}-\delta\Gamma^{\,\,\rho}_{\mu\,\rho;\nu}-2S^\sigma_{\phantom{\sigma}\rho\nu}\delta\Gamma^{\,\,\rho}_{\mu\,\sigma},
\label{var1}
\end{equation}
which can be verified directly from Eq.~(\ref{Ric2}) using the fact that $\delta\Gamma^{\,\,\rho}_{\mu\,\nu}$ is a tensor~\cite{Sch1950}.\footnote{
The variation of the curvature tensor is given by
\begin{equation}
\delta R^\rho_{\phantom{\rho}\mu\sigma\nu}=\delta\Gamma^{\,\,\rho}_{\mu\,\nu;\sigma}-\delta\Gamma^{\,\,\rho}_{\mu\,\sigma;\nu}-2S^\kappa_{\phantom{\sigma}\sigma\nu}\delta\Gamma^{\,\,\rho}_{\mu\,\kappa}.
\label{varRiem}
\end{equation}
}
Therefore, we obtain
\begin{equation}
\delta S=\int d^4x\,{\sf g}^{\mu\nu}(\delta\Gamma^{\,\,\rho}_{\mu\,\nu;\rho}-\delta\Gamma^{\,\,\rho}_{\mu\,\rho;\nu}-2S^\sigma_{\phantom{\sigma}\rho\nu}\delta\Gamma^{\,\,\rho}_{\mu\,\sigma}).
\label{var2}
\end{equation}

The formula for the covariant derivative of a scalar density $\Im$ (such as $\sqrt{-g}$) results from the covariant constancy of the Levi-Civita pseudotensor density, $\epsilon^{\mu\nu\rho\sigma}_{\phantom{\mu\nu\rho\sigma};\kappa}=0$~\cite{Sch1950}:
\begin{equation}
\Im_{;\mu}=\Im_{,\mu}-\Gamma^{\,\,\nu}_{\nu\,\mu}\Im.
\label{den}
\end{equation}
Combining this formula with the Gau\ss\,\,theorem and the chain rule for the covariant differentiation of the product of a tensor and a scalar density, we derive the identity
\begin{equation}
\int d^4x(\sqrt{-g}V^\mu)_{;\mu}=2\int d^4x\sqrt{-g}S_\mu V^\mu.
\label{Gauss}
\end{equation}
Integrating Eq.~(\ref{var2}) by parts and using~(\ref{Gauss}) we obtain
\begin{eqnarray}
& & \delta S=\int d^4x(-{\sf g}^{\mu\nu}_{\phantom{\mu\nu};\rho}\delta\Gamma^{\,\,\rho}_{\mu\,\nu}+{\sf g}^{\mu\nu}_{\phantom{\mu\nu};\nu}\delta\Gamma^{\,\,\rho}_{\mu\,\rho}+2{\sf g}^{\mu\nu}S_\rho\delta\Gamma^{\,\,\rho}_{\mu\,\nu} \nonumber \\
& & -2{\sf g}^{\mu\nu}S_\nu\delta\Gamma^{\,\,\rho}_{\mu\,\rho}-2{\sf g}^{\mu\nu}S^\sigma_{\phantom{\sigma}\rho\nu}\delta\Gamma^{\,\,\rho}_{\mu\,\sigma}) \nonumber \\
& & =\int d^4x(-{\sf g}^{\mu\nu}_{\phantom{\mu\nu};\rho}+{\sf g}^{\mu\sigma}_{\phantom{\mu\sigma};\sigma}\delta^\nu_\rho+2{\sf g}^{\mu\nu}S_\rho-2{\sf g}^{\mu\sigma}S_\sigma\delta^\nu_\rho-2{\sf g}^{\mu\sigma}S^\nu_{\phantom{\nu}\rho\sigma})\delta\Gamma^{\,\,\rho}_{\mu\,\nu}.
\label{var3}
\end{eqnarray}
According to the principle of least action, the variation of the action~(\ref{var3}) vanishes for an arbitrary variation $\delta\Gamma^{\,\,\rho}_{\mu\,\nu}$, which gives
\begin{equation}
{\sf g}^{\mu\nu}_{\phantom{\mu\nu};\rho}-{\sf g}^{\mu\sigma}_{\phantom{\mu\sigma};\sigma}\delta^\nu_\rho-2{\sf g}^{\mu\nu}S_\rho+2{\sf g}^{\mu\sigma}S_\sigma\delta^\nu_\rho+2{\sf g}^{\mu\sigma}S^\nu_{\phantom{\nu}\rho\sigma}=0,
\label{var4}
\end{equation}
or equivalently
\begin{equation}
{\sf g}^{\mu\nu}_{\phantom{\mu\nu};\rho}-2{\sf g}^{\mu\nu}S_\rho+2{\sf g}^{\mu\sigma}\biggl(S^\nu_{\phantom{\nu}\rho\sigma}+\frac{1}{3}S_\sigma\delta^\nu_\rho\biggr)=0.
\label{var5}
\end{equation}
By means of Eqs.~(\ref{con4}) and~(\ref{den}) we find 
\begin{equation}
{\sf g}^{\mu\nu}_{\phantom{\mu\nu},\rho}+\,^\ast\Gamma^{\,\,\mu}_{\sigma\,\rho}{\sf g}^{\sigma\nu}+\,^\ast\Gamma^{\,\,\nu}_{\rho\,\sigma}{\sf g}^{\mu\sigma}-\frac{1}{2}(^\ast\Gamma^{\,\,\sigma}_{\rho\,\sigma}+\,^\ast\Gamma^{\,\,\sigma}_{\sigma\,\rho}){\sf g}^{\mu\nu}=0,
\label{field1}
\end{equation}
which is the desired field equation relating the metric to the connection~\cite{Sch1950,Sch1947b}.

By contracting Eq.~(\ref{field1}) once with respect to $(\mu,\rho)$, then with respect to $(\nu,\rho)$, and subtracting the resulting equations, we obtain
\begin{equation}
{\sf g}^{[\mu\nu]}_{\phantom{[\mu\nu]},\nu}+\frac{1}{2}{\sf g}^{(\mu\nu)}(^\ast\Gamma^{\,\,\rho}_{\rho\,\nu}-\,^\ast\Gamma^{\,\,\rho}_{\nu\,\rho})=0.
\label{field2}
\end{equation}
Eqs.~(\ref{con5}) and~(\ref{field2}) yield
\begin{equation}
{\sf g}^{[\mu\nu]}_{\phantom{[\mu\nu]},\nu}=0,
\label{field3}
\end{equation}
which can be regarded as the field equation instead of~(\ref{con5})~\cite{Sch1950,Sch1947b}.\footnote{
Eq.~(\ref{field3}) can be obtained more directly by contracting Eq.~(\ref{var4}) with respect to $(\mu,\rho)$.}
We write Eq.~(\ref{var5}) as
\begin{equation}
g^{\mu\nu}_{\phantom{\mu\nu};\rho}\sqrt{-g}+g^{\mu\nu}(\sqrt{-g})_{,\rho}-{\sf g}^{\mu\nu}\Gamma^{\,\,\sigma}_{\sigma\,\rho}-2{\sf g}^{\mu\nu}S_\rho+2{\sf g}^{\mu\sigma}\biggl(S^\nu_{\phantom{\nu}\rho\sigma}+\frac{1}{3}S_\sigma\delta^\nu_\rho\biggr)=0,
\label{field4}
\end{equation}
and multiply by $g_{\mu\nu}$ to obtain\footnote{
We use the identity ${\sf g}^{\mu\nu}g_{\mu\nu;\rho}=2(\sqrt{-g})_{,\rho}-2\sqrt{-g}\,\Gamma^{\,\,\nu}_{\nu\,\rho}$.}
\begin{equation}
(\sqrt{-g})_{,\rho}-\sqrt{-g}\,\Gamma^{\,\,\sigma}_{\sigma\,\rho}=\frac{8}{3}\sqrt{-g}S_\rho.
\label{field5}
\end{equation}
Substituting Eq.~(\ref{field5}) to~(\ref{field4}) brings the latter into an explicitly tensorial form
\begin{equation}
g^{\mu\nu}_{\phantom{\mu\nu};\rho}+2S^\nu_{\phantom{\nu}\rho\sigma}g^{\mu\sigma}+\frac{2}{3}S_\rho g^{\mu\nu}+\frac{2}{3}S_\sigma\delta^\nu_\rho g^{\mu\sigma}=0.
\label{field6}
\end{equation}
Lowering both indices and using definition~(\ref{con4}) turns Eq.~(\ref{field6}) into the final expression~\cite{Sch1950,Sch1947b,ES,Ein1948,Ein1950}:\footnote{
Eq.~(\ref{field7}) contracted with $g^{\mu\nu}$ yields $(\mbox{ln}\sqrt{-g})_{,\mu}=\,^\ast\Gamma^{\,\,\,\,\nu}_{(\mu\,\nu)}$.}
\begin{equation}
g_{\mu\nu,\rho}-\,^\ast\Gamma^{\,\,\sigma}_{\mu\,\rho}g_{\sigma\nu}-\,^\ast\Gamma^{\,\,\sigma}_{\rho\,\nu}g_{\mu\sigma}=0.
\label{field7}
\end{equation}
This system of 64 equations can be solved for 64 unknown connection coefficients in terms of 16 components of the metric tensor~\cite{Str,Ing,Kur3,Bos,EK1,EK2}.
Lastly, we substitute Eq.~(\ref{con4}) into~(\ref{cosm3}) to obtain~\cite{Sch1950,Sch1947b}
\begin{equation}
^\ast R_{\mu\nu}+\frac{2}{3}(S_{\mu,\nu}-S_{\nu,\mu})=-\Lambda g_{\mu\nu},
\label{cosm4}
\end{equation}
where the tensor $^\ast R_{\mu\nu}$ is composed of $^\ast\Gamma^{\,\,\kappa}_{\rho\,\sigma}$ the same way $R_{\mu\nu}$ is composed of $\Gamma^{\,\,\kappa}_{\rho\,\sigma}$.

Eqs.~(\ref{field3}),~(\ref{field7}) and~(\ref{cosm4}) constitute the complete set of the field equations for the nonsymmetric purely affine gravity in vacuum~\cite{Sch1950,Sch1947b}.
Their number, 84, equals the total number of the unknown functions $g_{\mu\nu}$, $^\ast\Gamma^{\,\,\rho}_{\mu\,\nu}$ and $S_\mu$.
These equations coincide with the field equations of Einstein and Straus~\cite{ES,Ein1948,Ein1950} generalized by the cosmological term~\cite{Pap}.\footnote{
Eq.~(\ref{cosm4}) can be split into the symmetric part
\begin{equation}
^\ast R_{(\mu\nu)}=-\Lambda g_{(\mu\nu)},
\label{ES1}
\end{equation}
and the antisymmetric part written as~\cite{ES,Ein1948,Ein1950,Sch1950}
\begin{equation}
^\ast R_{[\mu\nu,\rho]}=-\Lambda g_{[\mu\nu,\rho]},
\label{ES2}
\end{equation}
which does not contain $S_\mu$ explicitly.}
The metric tensor can be eliminated by combining Eqs.~(\ref{cosm3}) and~(\ref{field7}) into
\begin{equation}
R_{\mu\nu,\rho}-\,^\ast\Gamma^{\,\,\sigma}_{\mu\,\rho}R_{\sigma\nu}-\,^\ast\Gamma^{\,\,\sigma}_{\rho\,\nu}R_{\mu\sigma}=0,
\label{cosm5}
\end{equation}
which has no analogue in the Einstein--Straus theory~\cite{Sch1950}.
This equation contains only the original variables $\Gamma^{\,\,\rho}_{\mu\,\nu}$ and does not contain $\Lambda$ explicitly, although it does embody the Einstein field equations with the cosmological term. 
It also follows that Eq.~(\ref{field3}) is a consequence of the self-contained Eq.~(\ref{cosm5})~\cite{Sch1950}.

\section{The Equations of Motion}
\label{secMot}

We assume that, in the purely affine gravity, a particle moves on an autoparallel curve defined as the curve $x^\mu(\lambda)$ such that the vector $\frac{dx^\mu}{d\lambda}$ tangent to it at any point, when parallely translated to another point on this curve, coincides with the tangent vector there~\cite{Edd,Sch1950,Eis3,Sch1944c}.
Using Eq.~(\ref{con1}), this condition can be written as
\begin{equation}
\frac{dx^\mu}{d\lambda}-\Gamma^{\,\,\mu}_{\rho\,\nu}\frac{dx^\rho}{d\lambda}dx^\nu=C\biggl(\frac{dx^\mu}{d\lambda}+\frac{d^2x^\mu}{d\lambda^2}d\lambda\biggr),
\label{geo1}
\end{equation}
where the proportionality factor $C$ is some function of $\lambda$.
Eq.~(\ref{geo1}) can be written as
\begin{equation}
C\frac{d^2x^\mu}{d\lambda^2}+\Gamma^{\,\,\mu}_{\rho\,\nu}\frac{dx^\rho}{d\lambda}\frac{dx^\nu}{d\lambda}=\frac{1-C}{d\lambda}\frac{dx^\mu}{d\lambda},
\label{geo2}
\end{equation}
from which it follows that $C$ must differ from 1 by the order of $d\lambda$.
Therefore, in the first term on the left-hand side of Eq.~(\ref{geo2}) we can put $C=1$, and we denote $1-C$ by $\phi(\lambda)d\lambda$.
Eq.~(\ref{geo2}) reads~\cite{Sch1950}
\begin{equation}
\frac{d^2x^\mu}{d\lambda^2}+\Gamma^{\,\,\mu}_{\rho\,\nu}\frac{dx^\rho}{d\lambda}\frac{dx^\nu}{d\lambda}=\phi(\lambda)\frac{dx^\mu}{d\lambda}.
\label{geo3}
\end{equation}

If we replace the variable $\lambda$ by $s(\lambda)$, Eq.~(\ref{geo3}) becomes
\begin{equation}
\frac{d^2x^\mu}{ds^2}+\Gamma^{\,\,\mu}_{\rho\,\nu}\frac{dx^\rho}{ds}\frac{dx^\nu}{ds}=\frac{\phi s'-s''}{{s'}^2}\frac{dx^\mu}{ds}.
\label{geo4}
\end{equation}
Requiring $\phi s'-s''=0$, which has the general solution $s=\int^\lambda d\lambda\,\mbox{exp}[-\int^\lambda \phi(x)dx]$, we bring Eq.~(\ref{geo4}) into the standard form~\cite{Sch1950}:\footnote{
In general relativity, the connection components are the Christoffel symbols, $\Gamma^{\,\,\rho}_{\mu\,\nu}=\{^{\,\,\rho}_{\mu\,\nu}\}$, and autoparallel curves are geodesic.}
\begin{equation}
\frac{d^2x^\mu}{ds^2}+\Gamma^{\,\,\mu}_{\rho\,\nu}\frac{dx^\rho}{ds}\frac{dx^\nu}{ds}=0,
\label{geo5}
\end{equation} 
where the variable $s$ is the affine parameter and $ds$ measures the length of an infinitesimal section of a curve.
The autoparallel Eq.~(\ref{geo5}) is invariant under linear transformations $s\rightarrow as+b$ since the two lower limits of integration in the expression for $s(\lambda)$ are arbitrary. 

Only the symmetric part $\Gamma^{\,\,\,\,\rho}_{(\mu\,\nu)}$ of the connection enters the equation of motion~(\ref{geo5}) because of the symmetry of $\frac{dx^\nu}{ds}\frac{dx^\rho}{ds}$.
At any point, a certain transformation of the coordinates brings all the components $\Gamma^{\,\,\,\,\rho}_{(\mu\,\nu)}$ to zero by means of Eq.~(\ref{con3}), ensuring that the principle of equivalence is satisfied~\cite{Veb}.
Defining $u^\mu=\frac{dx^\mu}{ds}$ and multiplying Eq.~(\ref{geo5}) by $g_{(\sigma\mu)}u^\sigma$ leads to~\cite{Wym}
\begin{equation}
\frac{d}{ds}\biggl(g_{(\sigma\mu)}u^\sigma u^\mu\biggr)+\biggl[-\frac{d}{ds}\biggl(g_{(\nu\sigma)}u^\nu\biggr)+\Gamma^{\,\,\,\,\mu}_{(\rho\,\nu)}g_{(\sigma\mu)}u^\nu u^\rho\biggr]u^\sigma=0.
\label{comp1}
\end{equation}
Using Eqs.~(\ref{field7}) and~(\ref{geo5}), we can write the second term on the left-hand side as
\begin{eqnarray}
& & -\frac{d}{ds}\biggl(g_{(\nu\sigma)}u^\nu\biggr)=-\frac{du^\nu}{ds}g_{(\nu\sigma)}-u^\nu u^\mu(g_{(\nu\sigma)})_{,\mu} \nonumber \\
& & =\Gamma^{\,\,\,\,\nu}_{(\rho\,\mu)}u^\mu u^\rho g_{(\nu\sigma)}-u^\mu u^\nu[\,^\ast\Gamma^{\,\,\,\,\rho}_{\nu(\mu}g_{|\rho|\sigma)}+\,^\ast\Gamma^{\,\,\,\,\rho}_{(\mu|\sigma|}g_{\nu)\rho}].
\label{comp2}
\end{eqnarray}
Eq.~(\ref{comp1}) becomes
\begin{eqnarray}
& & \frac{d}{ds}\biggl(g_{(\sigma\mu)}u^\sigma u^\mu\biggr)+u^\mu u^\rho u^\sigma[2\Gamma^{\,\,\,\nu}_{\rho\,\mu}g_{(\nu\sigma)}-\,^\ast\Gamma^{\,\,\,\nu}_{\rho\,\mu}g_{\nu\sigma}-\,^\ast\Gamma^{\,\,\,\nu}_{\mu\sigma}g_{\rho\nu}] \nonumber \\
& & =\frac{d}{ds}\biggl(g_{(\mu\nu)}u^\mu u^\nu\biggr)-\frac{4}{3}u^\mu u^\nu u^\rho S_\mu g_{\nu\rho}=0.
\label{comp3}
\end{eqnarray}

The Ricci tensor~(\ref{Ric2}) is invariant under a $\lambda$-transformation~\cite{Ein1953}:
\begin{equation}
\Gamma^{\,\,\,\rho}_{\nu\,\mu}\rightarrow\Gamma^{\,\,\,\rho}_{\nu\,\mu}+\delta^\rho_\nu\lambda_{,\mu},
\label{comp4}
\end{equation}
and so is the tensor $g_{\mu\nu}$.\footnote{
The $\lambda$ in Eq.~(\ref{comp4}) is a scalar function of the coordinates, and should not be confused with the parameter $\lambda$ describing the curve in Eq.~(\ref{geo1}).}
Consequently, the field equations~(\ref{field7}) are also invariant which can easily be seen since the star-affinity does not change under such a transformation.
We can write Eq.~(\ref{comp3}) as
\begin{equation}
d\,\mbox{ln}[g_{(\mu\nu)}u^\mu u^\nu]-\frac{4}{3}S_\mu dx^\mu=0.
\label{comp5}
\end{equation}
Applying a transformation~(\ref{comp4}) with $\lambda=-\int S_\mu dx^\mu$ turns Eq.~(\ref{comp5}) into
\begin{equation}
g_{(\mu\nu)}u^\mu u^\nu=\mbox{const}.
\label{comp6}
\end{equation}
This conservation is a manifestation of the result, found by Eisenhart, that the symmetric part $\Gamma^{\,\,\,\,\rho}_{(\mu\,\nu)}$ of the connection is not the only one which is compatible with the metric $g_{(\mu\nu)}$~\cite{Sch1947b,Eis3,Eis1,Eis2}.
The constant in Eq.~(\ref{comp6}) can be set to unity by normalizing the affine parameter, from which it follows that the tensor $g_{(\mu\nu)}$ generalizes the symmetric metric tensor of general relativity,\footnote{
This is equivalent to Eddington's interpretation of the tensor $R_{(\mu\nu)}$ as the geometric metric tensor, $ds^2=R_{(\mu\nu)}dx^\mu dx^\nu$~\cite{Edd}.}
\begin{equation}
g_{(\mu\nu)}dx^\mu dx^\nu=ds^2.
\label{comp7}
\end{equation}

The interpretation of $g_{(\mu\nu)}$ as the geometric metric tensor is related to the postulate that a particle moves on a geodesic.
Schr\"{o}dinger, in his great book, {\em Space-Time Structure}, pointed out that there are four possible tensors
that can play the part of the corresponding tensorial entity describing a gravitational field in general relativity:
$g_{(\mu\nu)}$, ${\sf g}^{(\mu\nu)}$, ${\sf g}_{(\mu\nu)}$, and $g^{(\mu\nu)}$.\footnote{The covariant tensor density ${\sf g}_{\mu\nu}$ is defined as the tensor density reciprocal to the contravariant tensor density ${\sf g}^{\mu\nu}$:
${\sf g}_{\mu\rho}{\sf g}^{\nu\rho}={\sf g}_{\rho\mu}{\sf g}^{\rho\nu}=\delta_\mu^\nu$.}
All four possibilities are different, but coincide in the limit when all the tensors are symmetric.
The first case was assumed by Papapetrou~\cite{Pap,Wym,Bon1,Bon2,Van} and later by Moffat~\cite{Mof1979,Mof1987}.
The second case was assumed by Kur\c{s}uno\u{g}lu and H\'{e}ly after reading a physical meaning from the contracted Bianchi identities~\cite{Kur4,Hel1,Hel2,Tre,Bor,Jon,Ant1,ALM1,Shi1,ALM2}.
The fourth case was assumed by Lichnerowicz, who proved, by availing only of the field equations, that $g^{(\mu\nu)}$ enters the eikonal equation as the contravariant metric, both geometrically and physically~\cite{Lich}.
The question of which tensor represents the physical metric should be answered by the purely affine theory of gravitation and electromagnetism~\cite{FK1981,Niko} extended to the nonsymmetric ${\sf g}^{\mu\nu}$.
The corresponding equations of motion will be contained in the field equations derived from this Lagrangian~\cite{EIH,EI1,IW,Wal,EI2,IS}.

The only solutions of the field equations~(\ref{field1}) and~(\ref{field3}) that correspond to physical fields are those for which the fundamental tensor density ${\sf g}^{\mu\nu}$ has the Lorentzian signature $(+,-,-,-)$.
This requirement is guaranteed by the condition~(\ref{phys}).

\section{The Centrally Symmetric Solution in Vacuum}
\label{secSol}

The condition for a tensor field $g_{\mu\nu}$ to be spherically symmetric is that an arbitrary rotation about the center of symmetry turns the functions $g_{\mu\nu}$ of $x^\rho$ into $g_{\mu'\nu'}$ which are the same functions of $x^{\rho'}$.  
The general form of the centrally symmetric tensor $g_{\mu\nu}\neq g_{(\mu\nu)}$ in the spherical coordinates in the Einstein--Straus theory~\cite{ES,Ein1948,Ein1950} was found by Papapetrou~\cite{Pap}:
\begin{equation}
g_{\mu\nu}=\left[ \begin{array}{cccc}
\gamma & -w & 0 & 0\\
w & -\alpha & 0 & 0\\
0 & 0 & -\beta & r^2v\,\mbox{sin}\,\theta\\
0 & 0 & -r^2v\,\mbox{sin}\,\theta & -\beta\,\mbox{sin}^2\theta \end{array} \right],
\label{sol1}
\end{equation}
where $\gamma$, $\alpha$, $\beta$, $w$ and $v$ are five arbitrary functions of the radial distance $r$.\footnote{The spherically symmetric solution turns out to be static, as in general relativity.}
The same form~(\ref{sol1}) must hold for the purely affine gravity.
Without loss of generality we can set $\beta=r^2$.
This choice implies that the variable $r$ is defined in such a way that the surface area of a sphere with center at the origin of the coordinates is equal to $4\pi r^2$. 

The two special solutions of Papapetrou describe the cases ($w\neq0$, $v=0$) and ($w=0$, $v\neq0$), respectively~\cite{Pap}.
Here, we examine in terms of the purely affine gravity only the first, simpler case.
Eq.~(\ref{field3}) gives
\begin{equation}
\frac{w^2 r^4}{\alpha\gamma-w^2}=l^4,
\label{sol2}
\end{equation}
where the length $l$ is a constant of integration.
The linear algebraic equations~(\ref{field7}) yield, as in the Einstein--Straus theory, 19 non-vanishing components of the star-affinity~\cite{Pap}:
\begin{eqnarray}
& & ^\ast\Gamma^{\,\,0}_{0\,1}=\,^\ast\Gamma^{\,\,0}_{1\,0}=\frac{\gamma'}{2\gamma}+\frac{2w^2}{r\alpha\gamma}; \nonumber \\
& & ^\ast\Gamma^{\,\,1}_{0\,0}=\frac{\gamma'}{2\alpha}+\frac{4w^2}{r\alpha^2},\,\,^\ast\Gamma^{\,\,1}_{1\,1}=\frac{\alpha'}{2\alpha},\,\,^\ast\Gamma^{\,\,1}_{2\,2}=-\frac{r}{\alpha},\,\,^\ast\Gamma^{\,\,1}_{3\,3}=-\frac{r}{\alpha}\mbox{sin}^2\theta; \nonumber \\
& & ^\ast\Gamma^{\,\,2}_{1\,2}=\,^\ast\Gamma^{\,\,2}_{2\,1}=\frac{1}{r},\,\,^\ast\Gamma^{\,\,2}_{3\,3}=-\mbox{sin}\,\theta\,\mbox{cos}\,\theta; \nonumber \\
& & ^\ast\Gamma^{\,\,3}_{1\,3}=\,^\ast\Gamma^{\,\,3}_{3\,1}=\frac{1}{r},\,\,^\ast\Gamma^{\,\,3}_{2\,3}=\,^\ast\Gamma^{\,\,3}_{3\,2}=\mbox{cot}\,\theta; \nonumber \\
& & ^\ast\Gamma^{\,\,1}_{0\,1}=-\,^\ast\Gamma^{\,\,1}_{1\,0}=-\frac{2w}{r\alpha},\,\,^\ast\Gamma^{\,\,2}_{0\,2}=-\,^\ast\Gamma^{\,\,2}_{2\,0}=\,^\ast\Gamma^{\,\,3}_{0\,3}=-\,^\ast\Gamma^{\,\,3}_{3\,0}=\frac{w}{r\alpha},
\label{sol3}
\end{eqnarray}
where the prime denotes the differentiation with respect to $r$.
Eqs.~(\ref{con4}), (\ref{Ric2}) and~(\ref{cosm4}) give the non-vanishing components of the star-Ricci tensor:
\begin{eqnarray}
& & ^\ast R_{00}=\biggl(\frac{\gamma'}{2\alpha}+\frac{4w^2}{r\alpha^2}\biggr)'-\biggl(\frac{\gamma'}{2\alpha}+\frac{4w^2}{r\alpha^2}\biggr)\biggl(\frac{\gamma'}{2\gamma}+\frac{2w^2}{r\alpha\gamma}-\frac{\alpha'}{2\alpha}+\frac{2}{r}\biggr)+\frac{6w^2}{r^2\alpha^2}, \nonumber \\
& & ^\ast R_{11}=\frac{\alpha'}{r\alpha}-\biggl(\frac{\gamma'}{2\gamma}+\frac{2w^2}{r\alpha\gamma}\biggr)'+\biggl(\frac{\gamma'}{2\gamma}+\frac{2w^2}{r\alpha\gamma}\biggr)\biggl(\frac{\alpha'}{2\alpha}-\frac{\gamma'}{2\gamma}-\frac{2w^2}{r\alpha\gamma}\biggr)=\Lambda\alpha, \nonumber \\
& & ^\ast R_{22}=\frac{1}{\mbox{sin}^2\theta}\,^\ast R_{33}=1-\frac{1}{\alpha}+\frac{r\alpha'}{2\alpha^2}-\frac{r}{\alpha}\biggl(\frac{\gamma'}{2\gamma}+\frac{2w^2}{r\alpha\gamma}\biggr)=\Lambda r^2, \nonumber \\
& & ^\ast R_{01}=-\,^\ast R_{10}=-\biggl(\frac{2w}{r\alpha}\biggr)'-\frac{4w}{r^2\alpha}=\Lambda w-\frac{2}{3}S_0'.
\label{sol4}
\end{eqnarray}
From Eq.~(\ref{sol2}) combined with the second and third equation in~(\ref{sol4}) one finds~\cite{Pap}:
\begin{eqnarray}
& & \gamma=\xi^2\biggl(1-\frac{r_g}{r}-\frac{\Lambda r^2}{3}\biggr)\biggl(1+\frac{l^4}{r^4}\biggr), \label{gamma}\\
& & \alpha=\biggl(1-\frac{r_g}{r}-\frac{\Lambda r^2}{3}\biggr)^{-1}, \label{alpha}\\
& & w=\pm\frac{l^2}{r^2}, \label{w}
\end{eqnarray}
where the length $r_g$ (the Schwarzschild radius) and the nondimensional $\xi^2$ are constants of integration.
The condition $\gamma\rightarrow1$ as $r\rightarrow\infty$ requires $\xi^2=1$.
The first equation in~(\ref{sol4}) introduces nothing new.
The last equation in~(\ref{sol4}) gives the only non-vanishing component of the torsion vector,
\begin{equation}
S_0=\pm l^2\biggl(\frac{1}{r^3}-\frac{3r_g}{2r^4}-\frac{\Lambda}{2r}\biggr),
\label{sol5}
\end{equation}
where we set the integration constant to zero to satisfy $S_0\rightarrow0$ as $r\rightarrow\infty$.\footnote{The Einstein--Straus Eq.~(\ref{ES2}) containing the component $R_{01}$ is satisfied identically~\cite{Pap}.}
For $l=0$, we reproduce the Schwarzschild solution of general relativity with the cosmological constant.

The spherically symmetric solution in vacuum given by Eqs.~(\ref{gamma}), (\ref{alpha}) and~(\ref{w}) has $\sqrt{-g}=r^2\mbox{sin}\,\theta$, which is the value corresponding to flat spacetime or the Schwarzschild metric.
There is a coordinate singularity at $r=r_g$ and a physical singularity at $r=0$.
However, if we assume that the star-affinity is complex and Hermitian in its lower indices, $\bar{^\ast\Gamma}^{\,\,\rho}_{\mu\,\nu}=\,^\ast\Gamma^{\,\,\rho}_{\nu\,\mu}$, so are the tensors $g_{\mu\nu}$ and $R_{\mu\nu}$ due to Eqs.~(\ref{cosm3}) and~(\ref{field7})~\cite{Ein1925,Ein1945}.
Consequently, $w$ becomes imaginary and $l^4$ becomes negative~\cite{Mof1977a,Mof1977b}.
If we replace $l^2$ by $iL^2$, the line element corresponding to this solution is given by
\begin{equation}
ds^2=\biggl(1-\frac{r_g}{r}-\frac{\Lambda r^2}{3}\biggr)\biggl(1-\frac{L^4}{r^4}\biggr)dt^2-\biggl(1-\frac{r_g}{r}-\frac{\Lambda r^2}{3}\biggr)^{-1}dr^2-r^2(d\theta^2+\mbox{sin}^2\theta d\phi^2).
\label{sing1}
\end{equation}
Its physical structure is the same as in Moffat's theory of gravitation~\cite{Mof1979,Mof1987,Mof1977a,Mof1977b} modifying the Einstein--Straus theory.
If $L>r_g$ then the region $r<L$ (including the black hole horizon and central singularity) is excluded from physical space since the interval $ds^2$ becomes negative as $r$ goes below $L$.
It can be shown that the null surface $r=L$ acts like a hard sphere deflecting all particles trying to go inside~\cite{Mof1977a,Mof1977b}.
If $L\leq r_g$, the black hole horizon is accessible for particles, but the singular point $r=0$ is not.
Overall, in the purely affine gravity with a nonsymmetric Hermitian metric, matter in a collapsing star never reaches the central singularity.
This nonsingular solution can be maximally extended using a transformation of the coordinates similar to that of Kruskal~\cite{Kru,KM}.

We do not examine here the second Papapetrou's special case since it is generally more complicated~\cite{Wym,Bon1,Bon2,Van}.
We only mention that this case in the Einstein--Straus--Moffat framework leads to the solution which is nonsingular everywhere in space, including the origin $r=0$~\cite{CM1,CM2}. 
There also exists an exact solution of the nonsymmetric gravity theory depending on three coordinates~\cite{Ant2}.

\section{The Second Ricci Tensor}
\label{secRic}

The second Ricci tensor~(\ref{Ric4}) has the form of the curl of the contracted connection $\Gamma^{\,\,\nu}_{\nu\,\mu}$~\cite{Scho}.
Actually, this tensor can be represented as the curl of a vector if we use the identity in the second footnote in Sec.~\ref{secField}:
\begin{equation}
Q_{\mu\nu}=\Gamma_{\nu,\mu}-\Gamma_{\mu,\nu},
\label{Ric5}
\end{equation} 
where the Weyl vector $\Gamma_\mu$~\cite{HK,HLS} measures nonmetricity:
\begin{equation}
\Gamma_\mu=-\frac{1}{2}g^{\nu\rho}g_{\nu\rho;\mu}.
\label{Ric6}
\end{equation}
The second Ricci tensor is invariant under a $\lambda$-transformation~(\ref{comp4}).
Therefore, without breaking the $\lambda$-invariance, we can embody $Q_{\mu\nu}$ into the gravitational Lagrangian density.
We cannot simply replace the Ricci tensor $R_{\mu\nu}$ by $Q_{\mu\nu}$ in Eq.~(\ref{Lagr1}) because the fundamental tensor density~(\ref{met1}) would turn out to be antisymmetric, and we could not construct any metric structure associated with such a Lagrangian.
Instead, we can assume that the modified Lagrangian density is the sum of two terms,
\begin{equation}
{\cal L}=-\frac{2}{\Lambda}\sqrt{-\mbox{det}R_{\mu\nu}}-\frac{2\alpha}{\Lambda}\sqrt{-\mbox{det}Q_{\mu\nu}},
\label{Lagr2}
\end{equation}
where $\alpha$ is a nondimensional constant.
Since both Ricci tensors $R_{\mu\nu}$ and $Q_{\mu\nu}$ are independent quantities,\footnote{If the connection is symmetric, then $Q_{\mu\nu}=2R_{[\mu\nu]}$ and it is sufficient to vary the action with respect to $R_{\mu\nu}$ only.}
the variation of the action becomes 
\begin{equation}
\delta S=\int d^4x\biggl(\frac{\partial{\cal L}}{\partial R_{\mu\nu}}\delta R_{\mu\nu}+\frac{\partial{\cal L}}{\partial Q_{\mu\nu}}\delta Q_{\mu\nu}\biggr).
\label{act3}
\end{equation}
The fundamental tensor density ${\sf g}^{\mu\nu}$ is again given by formula~(\ref{met1}).
In addition, we can construct a tensor density associated with the tensor $Q_{\mu\nu}$:
\begin{equation}
{\sf h}^{\mu\nu}=\frac{\partial{\cal L}}{\partial Q_{\mu\nu}},
\label{met4}
\end{equation}
which is antisymmetric and, in general, unrelated to ${\sf g}^{\mu\nu}$.

The field equations corresponding to the action~(\ref{Lagr2}) contain Eqs.~(\ref{field3}),~(\ref{field7}) and~(\ref{cosm4}), which we obtained by equaling the first term on the right-hand side of Eq.~(\ref{act3}) to zero.
We now require that the sum of both terms there vanishes.
The variation of the second Ricci tensor is given by
\begin{equation}
\delta Q_{\mu\nu}=\delta\Gamma^{\,\,\rho}_{\rho\,\nu,\mu}-\delta\Gamma^{\,\,\rho}_{\rho\,\mu,\nu}.
\label{varF1}
\end{equation}
Consequently, we find
\begin{equation}
\int d^4x\,\frac{\partial{\cal L}}{\partial Q_{\mu\nu}}\delta Q_{\mu\nu}=\int d^4x\,{\sf h}^{\mu\nu}(\delta\Gamma^{\,\,\rho}_{\rho\,\nu,\mu}-\delta\Gamma^{\,\,\rho}_{\rho\,\mu,\nu})=2\int d^4x\,{\sf h}^{\nu\sigma}_{\phantom{\nu\sigma},\sigma}\delta^\mu_\rho\,\delta\Gamma^{\,\,\rho}_{\mu\,\nu}.
\label{varF2}
\end{equation}
Adding Eq.~(\ref{varF2}) to the variation~(\ref{var3}) and using the principle of least action yield
\begin{equation}
{\sf g}^{\mu\nu}_{\phantom{\mu\nu};\rho}-{\sf g}^{\mu\sigma}_{\phantom{\mu\sigma};\sigma}\delta^\nu_\rho-2{\sf g}^{\mu\nu}S_\rho+2{\sf g}^{\mu\sigma}S_\sigma\delta^\nu_\rho+2{\sf g}^{\mu\sigma}S^\nu_{\phantom{\nu}\rho\sigma}-2{\sf h}^{\nu\sigma}_{\phantom{\nu\sigma},\sigma}\delta^\mu_\rho=0.
\label{varF3}
\end{equation}
Similarly to the tensor $g^{\mu\nu}$, we can introduce the contravariant tensor corresponding to the tensor density ${\sf h}^{\mu\nu}$:
\begin{equation}
h^{\mu\nu}=\frac{{\sf h}^{\mu\nu}}{\sqrt{-\mbox{det}{\sf h}^{\rho\sigma}}},
\label{h1}
\end{equation}
and use it to define the corresponding covariant tensor:
\begin{equation}
h^{\mu\nu}h_{\rho\nu}=\delta^\mu_\rho.
\label{h2}
\end{equation}
Following the procedure in the last paragraph of Sec.~\ref{secLagr}, we find
\begin{equation}
Q_{\mu\nu}=-\frac{\Lambda}{\alpha}h_{\mu\nu}.
\label{h3}
\end{equation}
Eq.~(\ref{Ric5}) is equivalent to $Q_{[\mu\nu,\rho]}=0$, which gives
\begin{equation}
h_{[\mu\nu,\rho]}=0.
\label{h4}
\end{equation} 
From Eq.~(\ref{h4}) it follows that
\begin{equation}
{\sf h}^{\mu\nu}_{\phantom{\mu\nu},\nu}=0,
\label{h5}
\end{equation}
and the field equations~(\ref{varF3}) reproduce the field equations~(\ref{var4}).
Therefore, the additive term in the Lagrangian density~(\ref{Lagr2}) containing the second Ricci tensor $Q_{\mu\nu}$ does not affect the spacetime structure of the purely affine gravity.

Let us consider another possibility, where the gravitational Lagrangian density is proportional to the square root of the determinant of a linear combination of both Ricci tensors:
\begin{equation}
{\cal L}=-\frac{2}{\Lambda}\sqrt{-\mbox{det}(R_{\mu\nu}+\beta Q_{\mu\nu})},
\label{Lagr3}
\end{equation}
where $\beta$ is a nondimensional constant.
In this case, the tensor densities ${\sf g}^{\mu\nu}$ and ${\sf h}^{\mu\nu}$, defined the same way as previously, are not independent since
\begin{equation}
{\sf h}^{\mu\nu}=\beta{\sf g}^{[\mu\nu]}.
\label{h6}
\end{equation}
Consequently, Eq.~(\ref{var4}) reads
\begin{equation}
{\sf g}^{\mu\nu}_{\phantom{\mu\nu};\rho}-{\sf g}^{\mu\sigma}_{\phantom{\mu\sigma};\sigma}\delta^\nu_\rho-2{\sf g}^{\mu\nu}S_\rho+2{\sf g}^{\mu\sigma}S_\sigma\delta^\nu_\rho+2{\sf g}^{\mu\sigma}S^\nu_{\phantom{\nu}\rho\sigma}-2\beta{\sf g}^{[\nu\sigma]}_{\phantom{[\nu\sigma]},\sigma}\delta^\mu_\rho=0,
\label{varF4}
\end{equation}
and Eq.~(\ref{field2}) becomes
\begin{equation}
(1+4\beta){\sf g}^{[\mu\nu]}_{\phantom{[\mu\nu]},\nu}+\frac{1}{2}{\sf g}^{(\mu\nu)}(^\ast\Gamma^{\,\,\rho}_{\rho\,\nu}-\,^\ast\Gamma^{\,\,\rho}_{\nu\,\rho})=0.
\label{varF5}
\end{equation}
Assuming $\beta\neq-\frac{1}{4}$\footnote{We can regard the singular case $\beta=-\frac{1}{4}$ as the limit $\beta\rightarrow-\frac{1}{4}$, so that the obtained general results are valid also for this case.} and using~(\ref{con5}) leads again to Eq.~(\ref{field3}).
As a result, the field equations~(\ref{field7}) remain unchanged.

Instead of Eqs.~(\ref{cosm3}) and~(\ref{h3}), we obtain
\begin{equation}
R_{\mu\nu}+\beta Q_{\mu\nu}=-\Lambda g_{\mu\nu},
\label{h7}
\end{equation}
which is equivalent to
\begin{equation}
^\ast R_{\mu\nu}+\frac{2}{3}(S_{\mu,\nu}-S_{\nu,\mu})+\beta(\Gamma_{\nu,\mu}-\Gamma_{\mu,\nu})=-\Lambda g_{\mu\nu}.
\label{h8}
\end{equation}
Therefore, the effect of the second Ricci tensor in the Lagrangian density~(\ref{Lagr3}) is simply to shift the torsion vector $S_\mu$ by a vector proportional to the nonmetricity vector $\Gamma_\mu$~(\ref{Ric6}).
The term with $Q_{\mu\nu}$ which contains the free parameter $\beta$ can be used to put one constraint on the components of the torsion vector $S_\mu$,
as does the gauge transformation for the electromagnetic potential.
The same effect can be achieved by applying a $\lambda$-transformation to the connection.
Consequently, the addition of the product of the second Ricci tensor $Q_{\mu\nu}$ and a constant scalar to the Ricci tensor $R_{\mu\nu}$ in the Lagrangian density~(\ref{Lagr1}) is physically equivalent to a $\lambda$-transformation involving a scalar function of the coordinates.

The purely affine formulation of gravity allows an elegant unification of the classical free electromagnetic and gravitational fields.
Ferraris and Kijowski showed that the gravitational field is represented by the symmetric part of the Ricci tensor $R_{(\mu\nu)}$ of the connection (not restricted to be symmetric), while the electromagnetic field can be represented by the second Ricci tensor $Q_{\mu\nu}$~\cite{FK1982b,Chr}.\footnote{Both $Q_{\mu\nu}$ and the electromagnetic field tensor $F_{\mu\nu}$ are curls.}
The purely affine Lagrangian density for the unified electromagnetic and gravitational fields is given by
\begin{equation}
{\cal L}=-\frac{e^2}{4}\sqrt{-\mbox{det}R_{(\mu\nu)}}Q_{\alpha\beta}Q_{\rho\sigma}P^{\alpha\rho}P^{\beta\sigma},
\label{FK}
\end{equation}
where $e$ has the dimension of electric charge and the tensor $P^{\mu\nu}$ is reciprocal to $R_{(\mu\nu)}$.
Since the Lagrangian density~(\ref{FK}) does not depend on $R_{[\mu\nu]}$, $\frac{\partial{\cal L}}{\partial R_{\mu\nu}}=\frac{\partial{\cal L}}{\partial R_{(\mu\nu)}}$ and the tensor density ${\sf g}^{\mu\nu}$ is symmetric.
This construction is dynamically equivalent to the sourceless Einstein--Maxwell equations~\cite{FK1982b,Chr}.

\section{Torsion}
\label{secTor}

A Lagrangian density is covariant if it is a product of a scalar and the square root of the determinant of a covariant tensor of rank two~\cite{Sch1950}.
The most general covariant tensor of rank two, constructed from the connection through the curvature and torsion tensors (and their derivatives), is a linear combination of the tensors:
$R_{(\mu\nu)}$, $R_{[\mu\nu]}$, $Q_{\mu\nu}$, $S^\rho_{\phantom{\rho}\mu\nu}S_\rho$, $S^\rho_{\phantom{\rho}\mu\nu;\rho}$, $S_\mu S_\nu$, $S_{(\mu;\nu)}$, and $S_{[\mu;\nu]}$~\cite{Shi2}, while there are no scalars constructed this way.\footnote{From the connection, one can construct scalars using tensors reciprocal to the contracted curvature tensors, e.g., $R_{(\mu\nu)}K^{\mu\nu}$.
Similarly, one can construct an infinite number of covariant second-rank tensors.}
In Sec.~\ref{secLagr} we assumed the covariant second-rank tensor in the gravitational Lagrangian to be the Ricci tensor.
In principle, any linear combination $\alpha R_{(\mu\nu)}+\beta R_{[\mu\nu]}$ could replace $R_{\mu\nu}$.
We will not study this case in this work, since, as it can be easily shown, we cannot arrive at a simple relation between $R_{\mu\nu}$ and $g_{\mu\nu}$ (as in Eq.~(\ref{cosm3})) using the definitions~(\ref{met1}) and~(\ref{met2}).\footnote{The case $\beta=0$ is the exception; it leads to Eq.~(\ref{cosm3}) with the symmetric $g_{\mu\nu}$, i.e. general relativity with the cosmological constant.}
In Sec.~\ref{secRic} we showed that including the second Ricci tensor $Q_{\mu\nu}$ in the purely affine Lagrangian does not affect the field equations and the equations of motion.

Let us consider the Lagrangian in which we add to the Ricci tensor a linear combination of the tensors: $S^\rho_{\phantom{\rho}\mu\nu}S_\rho$, $S^\rho_{\phantom{\rho}\mu\nu;\rho}$, $S_\mu S_\nu$, $S_{(\mu;\nu)}$, and $S_{[\mu;\nu]}$:
\begin{equation}
{\cal L}=-\frac{2}{\Lambda}\sqrt{-\mbox{det}(R_{\mu\nu}+aS^\rho_{\phantom{\rho}\mu\nu}S_\rho+bS^\rho_{\phantom{\rho}\mu\nu;\rho}+cS_\mu S_\nu+dS_{(\mu;\nu)}+eS_{[\mu;\nu]})},
\label{tors1}
\end{equation}
where $a$, $b$, $c$, $d$ and $e$ are nondimensional constants.
The variation of the action becomes 
\begin{eqnarray}
& & \delta S=\int d^4x\biggl(\frac{\partial{\cal L}}{\partial R_{\mu\nu}}\delta R_{\mu\nu}+\frac{\partial{\cal L}}{\partial(S^\rho_{\phantom{\rho}\mu\nu}S_\rho)}\delta(S^\rho_{\phantom{\rho}\mu\nu}S_\rho)}+\frac{\partial{\cal L}}{\partial S^\rho_{\phantom{\rho}\mu\nu;\rho}}\delta S^\rho_{\phantom{\rho}\mu\nu;\rho \nonumber \\
& & +\frac{\partial{\cal L}}{\partial(S_\mu S_\nu)}\delta(S_\mu S_\nu)+\frac{\partial{\cal L}}{\partial S_{(\mu;\nu)}}\delta S_{(\mu;\nu)}+\frac{\partial{\cal L}}{\partial S_{[\mu;\nu]}}\delta S_{[\mu;\nu]}\biggr) \nonumber \\
& & =\int d^4x\biggl({\sf g}^{\mu\nu}\delta R_{\mu\nu}+a{\sf g}^{[\mu\nu]}\delta(S^\rho_{\phantom{\rho}\mu\nu}S_\rho)}+b{\sf g}^{[\mu\nu]}\delta S^\rho_{\phantom{\rho}\mu\nu;\rho \nonumber \\
& & +c{\sf g}^{(\mu\nu)}\delta(S_\mu S_\nu)+d{\sf g}^{(\mu\nu)}\delta S_{(\mu;\nu)}+e{\sf g}^{[\mu\nu]}\delta S_{[\mu;\nu]}\biggr) \nonumber \\
& & = \int d^4x\biggl[-{\sf g}^{\mu\nu}_{\phantom{\mu\nu};\rho}+{\sf g}^{\mu\sigma}_{\phantom{\mu\sigma};\sigma}\delta^\nu_\rho+2{\sf g}^{\mu\nu}S_\rho-2{\sf g}^{\mu\sigma}S_\sigma\delta^\nu_\rho-2{\sf g}^{\mu\sigma}S^\nu_{\phantom{\nu}\rho\sigma} \nonumber \\
& & +a({\sf g}^{[\mu\nu]}S_\rho+{\sf g}^{\alpha\beta}S^{[\mu}_{\phantom{[\mu}\alpha\beta}\delta^{\nu]}_\rho)+b(2{\sf g}^{[\mu\nu]}S_\rho-{\sf g}^{[\mu\nu]}_{\phantom{[\mu\nu]};\rho}) \nonumber \\
& & +c({\sf g}^{(\mu\sigma)}S_\sigma\delta^\nu_\rho-{\sf g}^{(\nu\sigma)}S_\sigma\delta^\mu_\rho) \nonumber \\
& & +d\biggl({\sf g}^{(\mu\sigma)}S_\sigma\delta^\nu_\rho-{\sf g}^{(\nu\sigma)}S_\sigma\delta^\mu_\rho-\frac{1}{2}{\sf g}^{(\mu\sigma)}_{\phantom{(\mu\sigma)};\sigma}\delta^\nu_\rho+\frac{1}{2}{\sf g}^{(\nu\sigma)}_{\phantom{(\mu\sigma)};\sigma}\delta^\mu_\rho\biggr) \nonumber \\
& & +e\biggl({\sf g}^{[\mu\sigma]}S_\sigma\delta^\nu_\rho-{\sf g}^{[\nu\sigma]}S_\sigma\delta^\mu_\rho-\frac{1}{2}{\sf g}^{[\mu\sigma]}_{\phantom{(\mu\sigma)};\sigma}\delta^\nu_\rho+\frac{1}{2}{\sf g}^{[\nu\sigma]}_{\phantom{(\mu\sigma)};\sigma}\delta^\mu_\rho\biggr)\biggr]\delta\Gamma^{\,\,\rho}_{\mu\,\nu}.
\label{tors2}
\end{eqnarray}
Consequently, Eq.~(\ref{var4}) becomes
\begin{eqnarray}
& & {\sf g}^{\mu\nu}_{\phantom{\mu\nu};\rho}-{\sf g}^{\mu\sigma}_{\phantom{\mu\sigma};\sigma}\delta^\nu_\rho-2{\sf g}^{\mu\nu}S_\rho+2{\sf g}^{\mu\sigma}S_\sigma\delta^\nu_\rho+2{\sf g}^{\mu\sigma}S^\nu_{\phantom{\nu}\rho\sigma} \nonumber \\
& & -a({\sf g}^{[\mu\nu]}S_\rho+{\sf g}^{\alpha\beta}S^{[\mu}_{\phantom{[\mu}\alpha\beta}\delta^{\nu]}_\rho)-b(2{\sf g}^{[\mu\nu]}S_\rho-{\sf g}^{[\mu\nu]}_{\phantom{[\mu\nu]};\rho}) \nonumber \\
& & -c({\sf g}^{(\mu\sigma)}S_\sigma\delta^\nu_\rho-{\sf g}^{(\nu\sigma)}S_\sigma\delta^\mu_\rho) \nonumber \\
& & -d\biggl({\sf g}^{(\mu\sigma)}S_\sigma\delta^\nu_\rho-{\sf g}^{(\nu\sigma)}S_\sigma\delta^\mu_\rho-\frac{1}{2}{\sf g}^{(\mu\sigma)}_{\phantom{(\mu\sigma)};\sigma}\delta^\nu_\rho+\frac{1}{2}{\sf g}^{(\nu\sigma)}_{\phantom{(\mu\sigma)};\sigma}\delta^\mu_\rho\biggr) \nonumber \\
& & -e\biggl({\sf g}^{[\mu\sigma]}S_\sigma\delta^\nu_\rho-{\sf g}^{[\nu\sigma]}S_\sigma\delta^\mu_\rho-\frac{1}{2}{\sf g}^{[\mu\sigma]}_{\phantom{(\mu\sigma)};\sigma}\delta^\nu_\rho+\frac{1}{2}{\sf g}^{[\nu\sigma]}_{\phantom{(\mu\sigma)};\sigma}\delta^\mu_\rho\biggr)=0.
\label{tors3}
\end{eqnarray}
By contracting Eq.~(\ref{tors3}) with respect to $(\mu,\rho)$ and $(\nu,\rho)$ we obtain, respectively:
\begin{eqnarray}
& & \biggl(1+\frac{b}{2}+\frac{3e}{4}\biggr){\sf g}^{[\mu\sigma]}_{\phantom{[\mu\sigma]};\sigma}+\frac{3d}{4}{\sf g}^{(\mu\sigma)}_{\phantom{(\mu\sigma)};\sigma}-\biggl(1+\frac{3a}{4}\biggr){\sf g}^{\rho\nu}S^\mu_{\phantom{\mu}\rho\nu}-\frac{3}{2}(c+d){\sf g}^{(\mu\sigma)}S_\sigma \nonumber \\
& & -\biggl(2+\frac{a}{2}+b+\frac{3e}{2}\biggr){\sf g}^{[\mu\sigma]}S_\sigma=0,
\label{tors4}
\end{eqnarray}
\begin{eqnarray}
& & \biggl(1-\frac{b}{3}-\frac{e}{2}\biggr){\sf g}^{[\mu\sigma]}_{\phantom{[\mu\sigma]};\sigma}+\biggl(1-\frac{d}{2}\biggr){\sf g}^{(\mu\sigma)}_{\phantom{(\mu\sigma)};\sigma}+\frac{a}{2}{\sf g}^{\rho\nu}S^\mu_{\phantom{\mu}\rho\nu}+\biggl(-\frac{4}{3}+c+d\biggr){\sf g}^{(\mu\sigma)}S_\sigma \nonumber \\
& & +\biggl(-\frac{4}{3}+\frac{a}{3}+\frac{2b}{3}+e\biggr){\sf g}^{[\mu\sigma]}S_\sigma=0.
\label{tors5}
\end{eqnarray}
Eqs.~(\ref{tors4}) and~(\ref{tors5}) give
\begin{eqnarray}
& & \biggl(1+\frac{b}{2}-\frac{5d}{4}+\frac{3e}{4}\biggr){\sf g}^{[\mu\sigma]}_{\phantom{[\mu\sigma]},\sigma}-\frac{1}{2}(3c+d){\sf g}^{(\mu\sigma)}S_\sigma \nonumber \\
& & +\frac{1}{2}\biggl(b+\frac{3}{2}(e-a-d)\biggr){\sf g}^{\rho\nu}S^\mu_{\phantom{\mu}\rho\nu}-\frac{1}{2}(a+d){\sf g}^{[\mu\sigma]}S_\sigma=0.
\label{tors6}
\end{eqnarray}

Eq.~(\ref{field3}) remains valid if $c=\frac{a}{3}$, $d=-a$ and $e=-\frac{2b}{3}$, assuming $a\neq-\frac{4}{5}$.\footnote{We can regard the singular case $a=-\frac{4}{5}$ as the limit $a\rightarrow-\frac{4}{5}$, so that the obtained general results are valid also for this case.}
To obtain the same relation between the fundamental tensor density and the star-affinity as in formula~(\ref{field1}), we need all the terms on the left-hand side of Eq.~(\ref{tors3}) except the first five to equal to the expression of form $A{\sf g}^{[\mu\sigma]}_{\phantom{[\mu\sigma]},\sigma}\delta^\nu_\rho+B{\sf g}^{[\nu\sigma]}_{\phantom{[\nu\sigma]},\sigma}\delta^\mu_\rho$, where $A$ and $B$ are constants.
This condition is impossible unless $a=b=0$, which, assuming the validity of Eq.~(\ref{field3}), yields $c=d=e=0$.
Therefore, the Lagrangian~(\ref{tors1}) is dynamically inequivalent to the Lagrangian~(\ref{Lagr1}).
Torsion inside the determinant in the Lagrangian, unlike the second Ricci tensor, affects the dynamics of the gravitational field in the purely affine theory of gravity.

\section{Summary}
\label{secSum}

The main objective of this paper was to review the vacuum purely affine gravity with the nonsymmetric connection and metric, and examine how the second Ricci tensor and covariant second-rank tensors built of the torsion tensor affect the dynamics of the gravitational field in vacuum in the nonsymmetric purely affine theory of gravity.
We found that the addition of the second Ricci tensor to the Ricci tensor inside the determinant in the Lagrangian density of the purely affine gravitational field is physically equivalent to the $\lambda$-transformation of the connection, i.e. the dynamics of the gravitational field does not depend on the second Ricci tensor.
However, the presence of the torsion tensor in the Lagrangian does change the dynamics of the gravitational field, as it occurs in the metric formulation of gravity with torsion.
We restricted our analysis to Lagrangians that do not contain tensors reciprocal to the Ricci tensors. We also did not explore the case where the second-rank tensor in the gravitational Lagrangian is a linear combination of the symmetric and antisymmetric part of the Ricci tensor.
This case does not give a simple relation between the metric and curvature tensors.

\section*{Acknowledgment}

The author wishes to thank Salvatore Antoci for valuable comments.

\end{document}